\documentclass[fleqn,usenatbib]{mnras}

\usepackage{newtxtext,newtxmath}
\usepackage[T1]{fontenc}

\DeclareRobustCommand{\VAN}[3]{#2}
\let\VANthebibliography\thebibliography
\def\thebibliography{\DeclareRobustCommand{\VAN}[3]{##3}\VANthebibliography}

\def\w{$\omega$}
\def\noise{$\epsilon$}
\def\chandra{{\em Chandra}}
\def\gaia{{\em Gaia}}
\def\erosita{{\em eROSITA}}
\def\msun{M$_\odot$}

\def\mbh{$M_{\rm BH}$}

\def\p{$\pm$}
\def\nh{$N_{\rm H}$}

\def\bprp{$B_{\rm p}$\,--\,$R_{\rm p}$}

\def\ltsim{\mathrel{\hbox{\rlap{\hbox{\lower4pt\hbox{$\sim$}}}\hbox{$<$}}}}
\def\gtsim{\mathrel{\hbox{\rlap{\hbox{\lower4pt\hbox{$\sim$}}}\hbox{$>$}}}}

\def\pertenmille{\ensuremath{{}^\text{o}\mkern-5mu/\mkern-3mu_\text{ooo}}}

\usepackage{graphicx}	% Including figure files

\usepackage{amsmath}	% Advanced maths commands
\usepackage{amssymb}	% Extra maths symbols

\title[X-ray sources with astrometric excess noise]{Astrometric excess noise in Gaia EDR3 and the search for X-ray binaries}
\author[P. Gandhi et al.]{P.\,Gandhi,$^{1,\,2}$\thanks{E-mail: poshak.gandhi@soton.ac.uk (PG)} 
D.A.H.\,Buckley,$^{3}$
P.A.\,Charles,$^1$
S.\,Hodgkin,$^{4}$
S.\,Scaringi,$^{5}$
C.\,Knigge,$^1$
A.\,Rao,$^{1,6}$
\newauthor
J.A.\,Paice,$^1$
Y.\,Zhao$^{7,1}$
\\
$^{1}$School of Physics \& Astronomy, University of Southampton, Highfield SO17 1BJ\\
$^{2}$Inter-University Centre for Astronomy \& Astrophysics, Post Bag 4, Ganeshkhind, Pune, 411007, India\\
$^{3}$South African Astronomical Observatory, PO Box 9, Observatory 7935, Cape Town, South Africa\\
$^{4}$Institute of Astronomy, Madingley Road, University of Cambridge, Cambridge, CB3 0HA\\
$^{5}$Department of Physics, Durham University, South Road, Durham, DH1 3LE\\
$^{6}$Department of Astronomy, Astrophysics \& Space Engineering, Indian Institute of Technology, Indore, Khandwa Road, Simrol, Indore 453552, India\\
$^{7}$Department of Physics, University of Alberta, 4--181 CCIS, Edmonton, Alberta, Canada T6G 2E1\\
}

\date{Accepted 2021 Dec 16; Submitted 2020 Aug 27; Revised 2021 Nov 18}

\pubyear{2021}

\begin{document}
\label{firstpage}
\pagerange{\pageref{firstpage}--\pageref{lastpage}}
\maketitle

\begin{abstract}
Astrometric noise (\noise) in excess of parallax and proper motion is a potential signature of orbital wobble (\w) of individual components in binary star systems. The combination of X-ray selection with astrometric noise could then be a powerful tool for robustly isolating accreting binaries in large surveys. Here, we mine the \gaia\ EDR3 catalogue for Galactic sources with significant values of astrometric noise over the parameter space expected for known and candidate X-ray binaries (XRBs). Cross-matching our sample with the \chandra\ Source Catalogue returns a primary sample of $\approx$\,6,500 X-ray sources with significant \noise. X-ray detection efficiency for objects with significant \noise\ is a factor of $\approx$\,4.5\,times higher than in a matched control sample exhibiting low \noise. The primary sample branches off the main sequence much more than control objects in colour-mag space, and includes a higher fraction of known binaries, variables and young stellar object class types. However, values of \noise\ reported in the \gaia\ pipeline releases so far can exceed expectations for individual XRBs with known semi-major axis size and other system parameters. It is likely that other factors (possibly attitude and modelling uncertainties, as well as source variability) currently dominate the observed excess noise in such systems. Confirmation of their nature must therefore await future \gaia{} releases. The full X-ray matched catalogue is released here to enable legacy follow-up.  
\end{abstract}


%%%%%%%%%%%%%%%%%%%%%%%%%%%%%%%%%%%%%%%%%%%%%%%%%%

%%%%%%%%%%%%%%%%% BODY OF PAPER %%%%%%%%%%%%%%%%%%

\section{Introduction}

Understanding final stage binary evolution is key to tracing the life cycle of the population of interacting binaries. There has been a surge in interest in compact binaries following the discovery of gravitational wave sources \citep{ligo1}. But it remains unclear if and how the current LIGO/Virgo populations connects to the electromagnetically observed accreting binary sources, e.g. do they cover systematically different parameter space in mass, natal kicks, compact object spin, and evolutionary history, to name just a few \citep[e.g. ][]{mandel16, gandhi20, jonker21, fishbach21}?

A key bottleneck here is the paltry number of confirmed stellar-mass black holes. There are only about 25 known stellar-mass black holes in the Milky Way with dynamical mass estimates. All of them lie in binary systems where spectroscopic radial velocity variations of the companion star have been used to confirm the presence of massive compact objects \citep{blackcat, watchdog}. By contrast, the Galaxy is expected to host anywhere between $\sim$\,10$^3$--10$^8$ stellar-mass black holes in binary systems \citep[e.g. ][]{pfahl03,tetarenko16}. This population should be dominated by non-accreting systems and incipient black hole X-ray binaries (BHXBs) with only a handful of recent well studied systems \citep{thompson19, tetarenko16, rivinius20}. An interesting recent highlight in this field was LB-1, with a proposed \mbh\,=\,70\,\msun\ \citep{liu19_lb1}. Though now believed to be much lighter \citep[e.g. ][]{eldridge20, elbadry21_hr6819, Abdul-Masih20, irrgang20}, its discovery accelerated efforts to understand the space density of massive quiescent BHXBs. Other examples include the report of a putative black hole in V723\,Mon \citep{Jayasinghe21}, in the triple system HR\,6819 \citep{rivinius20}, and in the open cluster NGC\,1850 \citep{saracino21}, though the latter two also remains controversial \citep{bodensteiner20, elbadryburdge21}. Clearly, more efforts are needed to enhance this population if we are to properly constrain final stage binary evolution. 

The exquisite astrometric precision now being enabled by missions such as \gaia\ opens up a new window on such studies. In particular, orbital motion of the companion star in a binary system will result in `astrometric orbital wobble (\w)', over and above the parallax and proper motion locus determined for single-object astrometry. This can manifest as an `astrometric excess noise (\noise)', one of the parameters reported in the early data release \citealt{gaiadr2, gaiaedr3}). \noise\ is defined as the excess uncertainty that must be added in quadrature to obtain a statistically acceptable astrometric solution \citep{gaiadr2, lindegren12}.\footnote{\url{https://gea.esac.esa.int/archive/documentation/GDR2/Gaia_archive/chap_datamodel/sec_dm_main_tables/ssec_dm_gaia_source.html}}. In the early data releases, \noise\ includes instrument and attitude modelling errors that are statistically significant and could result in large values of \noise. Thus, a detailed investigation of \noise-based selection is warranted, and this is what we carry out herein. 

Several recent theoretical works have highlighted the feasibility of large surveys, including astrometric missions  \citep{gould02, barstow14_gaia_wp, breivik17, mashian17, yalinewich18, andrews19, chawla21} and microlensing searches \citep{masuda19, wiktorowicz21}, to uncover large new populations of black holes in binary orbits. Massive spectroscopic surveys are also beginning to probe this territory through brute force blind searches for radial velocity variations characteristic of massive compact objects \citep{yi19_lamost, wiktorowicz20,  pricewhelan20}. Furthermore, large multiwavelength surveys such as \erosita\ in X-rays \citep{erosita} and ngVLA in the radio \citep{maccaronengvla} will be instrumental in confirming the nature of newly identified candidates, and characterising their physical properties (e.g. with the Rubin Observatory; \citealt{johnson19}). Thus, there are enormous synergies waiting to be explored in this field. We exploit one such synergy of astrometric noise combined with X-ray photometry here.

\section{Astrometric wobble and excess noise}
\label{sec:understanding}

Astrometric wobble (\w) is defined here simply as the maximal projected half angle swept by the companion star over its orbit, if the observed flux is dominated by the companion star. 

\begin{equation}
    \omega = \frac{a_2}{d}
    \label{eq:wobble}
\end{equation}

\noindent
where $a_2$ is the semi-major axis of the companion and $d$ the source distance. For the typical physical parameters and distances of known XRBs, \w\ is expected to lie in the range of $\sim$\,0.01--1.0\,mas (e.g. \citealt{gandhi19}).\footnote{the only difference being that $\omega$ was defined in \citet{gandhi19} to be the full swept angle over an orbit, twice the value defined here.} For simplicity, a circular orbit is assumed in these approximate estimates, since interacting binaries (our core targets of interest) are expected to circularise rapidly. While this is not the case for the (often highly) eccentric Be X-ray systems (see e.g. \citealt{Reig11}), virtually all known Be X-ray sources have neutron star (NS) compact objects, so the astrometric wobble of the (much) more massive Be donor is expected to be very small. The assumption of the flux being dominaed by the companion should be mostly true in quiescent, non-accreting, and high-mass XRBs, though the accreting primary in low-mass XRBs can contribute a few tens of per cent of the total flux (which would result in a smaller apparent centre-of-light wobble). 

Astrometric noise \noise\ represents an additional intrinsic scatter term in the \gaia\ pipeline astrometric solution, where it is expressed in angular units of mas. This is the value that needs to be added in quadrature to the formal statistical uncertainties in order to make the single-object solution statistically acceptable, effectively making the reduced sum-of-squared-weighted-residuals equal unity \citep{lindegren12}. 

Since $\omega$ represents a deviation from the nominal parallax locus, it is equivalent to an excess scatter term, and is thus conceptually similar to \noise, if no other noise terms contribute. In such a case, the expectation value of \noise\ should approximate $\hat{\omega}$\,=\,$\omega$/$\sqrt{2}$ in the limit of perfect orbital sampling. However, in early \gaia\ releases, \noise\ absorbs instrumental as well as attitude modelling errors that are likely to be statistically significant. The excess noise terms are globally adjusted to match the weighted sum of residuals to the number of degrees of freedom \citep{lindegren18}, so there could be some potential degeneracy between the magnitudes of the noise terms, and they need not scale directly and strictly with $\omega$ for individual objects. Thus, caution is needed in their interpretation \citep{lindegren12, luri18, gaiadr2, gaiaedr3}, and other supporting evidence should be leveraged as we describe in the following section. 

As an aside, astrometric perturbations to single-source pipeline fits are quantified in the \gaia\ pipeline in a variety of ways. In addition to \noise, the \gaia\ data releases include a parameter statistic termed RUWE (Renormalised Unit Weight Error). This is equivalent to a goodness-of-fit renormalised to 1 after accounting for systematic pipeline issues including a degrees-of-freedom bug as well as fit variations based upon colour \citep{lindegren18}. Significantly higher values than 1 can be signposts of intrinsic source complexity. RUWE selection and \noise\ selection are thus complementary to each other, with each having its pros and cons \citep[e.g., ][]{belokurov20,penoyre20}. Here, we are focusing on \noise\ due to its straightforward interpretation as a characteristic projected binary size. Another advantage is that the \gaia\ pipeline quantifies and reports the significance of \noise, unlike RUWE. The \gaia\ team have explored the regime where care is needed with \noise\ selection and we have adopted their recommendations (see following section). In any case, the aforementioned pipeline systematic issues will, at worst, result in an {\em underestimate} of \noise, so our selection is likely somewhat conservative.

\section{Sample selection}

\subsection{Mining \gaia}

With the reference parameter range of known XRBs discussed above as a starting ansatz, we used the following selection criteria. The \gaia\ EDR3 archive reports a {\tt significance} value of $\ge$\,2 when \noise\ is considered significant. At mags $G$\,$<$\,13, there are  systematic calibration uncertainties in EDR3, resulting in artificially enhanced values of \noise\ \citep{gaiaedr3,lindegren21}.\footnote{\citet{lindegren21} additionally suggest checks on the EDR3 effective wave number parameter $\nu_{\rm eff}$,  but we found this potentially impacts only a handful of objects.} Conversely, statistical uncertainties dominate near the faint mag limit of $G$\,=\,21. We thus restricted our mag range to 13\,$<$\,$G$\,$<$\,20. This encompasses the median mag of BHXBs with five-parameter astrometric solutions measured in DR2 ($G$\,$\approx$\,17.4; \citealt{gandhi19}), and should probe the mag range that is currently most robust to the aforementioned uncertainties. A distance range of 0.1--10\,kpc was examined, requiring a significant parallax (distance) measurement in order to try and assess the nature of the source population, as discussed in the following section. A minimum threshold on the number of visibility periods is used to ensure adequate sampling in time in the astrometric fit. The corresponding EDR3 ADQL \citep{adql} query for our primary sample over the full sky is:
~\newline

\noindent
{\tt
SELECT *\\
FROM gaiaedr3.gaia\_source\\
WHERE (astrometric\_excess\_noise >= 0.01) \hspace*{0.25cm}AND\\ 
\hspace*{0.25cm}(astrometric\_excess\_noise\_sig >= 2) \hspace*{0.25cm}AND\\
\hspace*{0.25cm}(parallax $<$ 10.) AND (parallax $>$ 0.1) AND\\ 
\hspace*{0.25cm}(parallax\_over\_error $>$ 5) AND \hspace*{0.25cm}(visibility\_periods\_used $>$ 10) AND\\
\hspace*{0.25cm}(phot\_g\_mean\_mag $>$ 13) AND\\
\hspace*{0.25cm}(phot\_g\_mean\_mag $<$ 20).
}
~\newline

\noindent
We also defined a control sample for cross-comparison. The EDR3 ADQL query for this control sample is identical to the above except for the excess noise selection criterion, because this is our main parameter of interest. We thus modify the relevant portion of the query with a complementary criterion, as follows:

{\tt
(astrometric\_excess\_noise < 0.01).
}
~\newline

%\noindent
%The above selection criteria are termed `G1', for the respective \gaia\ primary and control samples. 

Good astrometric fits require sources to be free from confusion and blending with close neighbours. Therefore, we next excluded all objects with any detected EDR3 near-neighbours.  %(criterion G2). 
A radius of 2\arcsec\ was adopted for our near-neighbour threshold, given that the nominal \gaia\ point spread function is concentrated well within 1\,arcsec.\footnote{\url{https://gea.esac.esa.int/archive/documentation/GDR2/Data_processing/chap_astpre/sec_cu3pre_cali/ssec_cu3pre_cali_psflsf.html}} This mitigates crowding issues impacting astrometry in dense regions of the Galactic plane. %This exclusion was applied to both the the primary and the control samples.

The final parallaxes that we report have been corrected for zero-point astrometric offsets, as recommended by the \gaia\ team \citep{lindegren21_zpoffset}; these were calculated from the mean photometeric $G$ band magnitude ({\tt phot\_g\_mean\_mag}), $\nu_{eff}$ ({\tt nu\_eff\_used\_in\_astrometry}), the pseudocolour ({\tt pseudocolour}), the ecliptic latitude ({\tt ecl\_lat}) and the number of astrometric parameters solved ({\tt astrometric\_params\_solved}). The resultant zeropoint offsets, calculated as described in \citet{lindegren21_zpoffset}, were then subtracted from the raw parallaxes reported by EDR3 pipeline.

\subsection{Cross-match with X-rays}

X-ray activity is a key signature of accretion in  binaries.  
Quiescent BHXBs are expected to exhibit low-level accretion activity, with typical luminosities of up to $L_{\rm X}$\,$\sim$\,10$^{30-32}$ erg\,s$^{-1}$, and NSXBs in quiescence can be even more luminous, on average \citep[e.g. ][]{reynoldsmiller11}. However, detection of X-rays by itself is not unambiguous proof of the presence of an interacting binary, with other possibilities including magnetically active stars \citep[e.g. ][]{Gudel04}, colliding winds \citep[e.g. ][]{PIttard18}, and activity in young stellar objects \citep[e.g. ][]{feigelson99}. So care will be necessary when making final inferences.

We queried the \chandra{} Source Catalogue (CSC; \citealt{csc}) for overlap with our sample. This is still one of the largest, and most sensitive, public databases in terms of broadband X-ray sky coverage, delivering exquisite spatial resolution ($\ltsim$\,1\arcsec\ on axis). High precision centroiding is critically important in crowded regions such as the Galactic plane, along which many of our sources will fall. The latest data release, CSC2.0 \citep{csc2},  covers approximately 550\,deg$^2$ (1.3\,\%) of the sky down to a point source sensitivity limit of 5\,counts. Assuming a reference spectrum of an accreting source characterised by an X-ray power-law with slope $\Gamma$\,=\,2,\footnote{Photon rate density $N(E)$\,$\propto$\,$E^{-\Gamma}$ at energy $E$.} this sensitivity corresponds to a 0.5--7\,keV flux limit $F_{\rm X}$ of 6\,$\times$\,10$^{-15}$\,erg\,cm$^{-2}$\,s$^{-1}$ at the median CSC2 exposure time of 12\,ks.\footnote{\url{https://cxc.harvard.edu/csc/char.html}} This is an approximation based upon the latest response function\footnote{\url{https://heasarc.gsfc.nasa.gov/cgi-bin/Tools/w3pimms/w3pimms.pl}} and assuming a line-of-sight column density \nh\,=\,5\,$\times$\,10$^{21}$\,cm$^{-2}$, not atypical out to distances of a few kpc in the Galactic plane. The \chandra\ soft energy response has been degrading with time so it is likely that older observations were more sensitive, on average. Taking the above flux limit as a baseline for comparison, CSC2 should be able to detect XRBs out to a distance $d$ with luminosity greater than 

\begin{equation}
    L_{\rm X-ray}>7\times 10^{29} \left(\frac{d}{1\,\rm{kpc}}\right)^{2}\,{\rm erg\,s^{-1}\,\,\, [0.5-7\, keV]}.
\end{equation}

\noindent
A maximal optical/X-ray cross-matching offset of 1$\arcsec$ was adopted, after back-tracing the \gaia\ 2016 reference epoch coordinates to 2000 using their EDR3 proper motions. We used the broad band (0.5--7\,keV) fluxes listed under the {\tt flux\_aper\_b} parameter. In a small fraction of cases, a non-zero value of the wide-band 0.1--10\,keV {\tt flux\_aper\_w} parameter is instead found, and this was converted to an equivalent broad-band flux. There is also a small fraction ($\sim$\,10\%) of sources where a flux measurement fails completely. We nevertheless retained these sources for some of the statistical analysis presented later, as their exclusion did not significantly impact our inferences.

\section{Results}
\label{sec:results}

\begin{table}
    \centering
    \begin{tabular}{lcr}
\hline
\hline
Selection basis & Primary & Control\\
\hline
\gaia\ EDR3 & 18,682,537 & 96,044,222 \\
\gaia\ + \chandra\ & 6,569 & 7,412 \\ 
\hline
    \end{tabular}
            \caption{Sample selection statistics
\label{tab:stats}}
\end{table}

Table\,\ref{tab:stats} lists the number of sources selected under various criteria. Our \gaia/EDR3 mining resulted in over 18 million sources selected in the `primary' sample. These are sources with significant astrometric excess noise and no close neighbours. The corresponding `control' sample was much larger as expected, approaching 100 million sources. The peak mags for primary and control are, respectively, $G$\,=\,15.7 and $G'$\,=\,16.4. Here, and hereafter, a prime ($'$) superscript refers to the control sample. The control sample objects tend to lie farther than primary sources, with mean distances of $\langle d'\rangle$\,=\,1.9\,kpc and  $\langle d\rangle$\,=\,1.2\,kpc, respectively, and a standard deviation of 1.0\,kpc for both. Here, we use parallax inversion $d$\,(kpc)\,=\,$\frac{1}{\pi}$, where $\pi$\ is the zero-point-corrected EDR3 parallax in mas. Parallax inversion should be a fair estimator of the distance if $\pi$ is well constrained, and certainly reasonable for population-wide comparisons. 

With this selection, the full \gaia-only selected sample ends up with a \noise\ value distribution peaking near \noise\,$\approx$\,0.25\,mas, but with an extended tail to $\approx$\,17\,mas. By contrast, the vast majority (99.9\,\%) of control sample objects have \noise$'$\,=\,0 mas. The above differences in characteristic brightness and distances between the primary and the control samples likely reflects the fact that detecting significant intrinsic perturbations to static single-object astrometric fits is simply more effective when sources are nearer and/or brighter. 

\begin{figure}
\hspace*{-0.7cm}
	\includegraphics[angle=90,width=1.1\columnwidth]{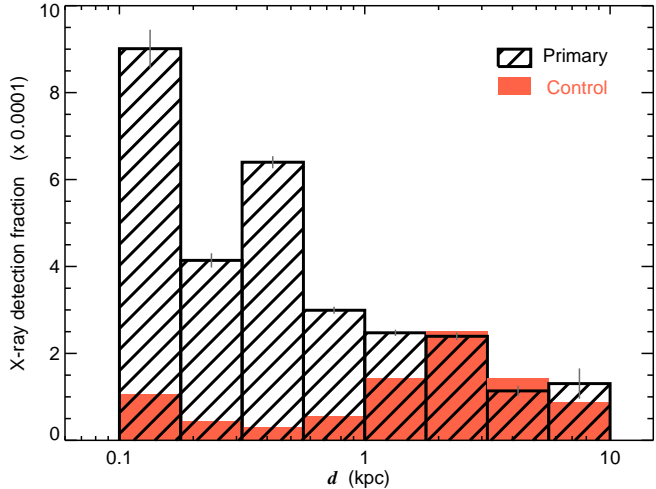}
	\caption{X-ray detection fraction ($f_X$) as a function of source distance $d$. $f_X$ is consistently larger in the primary sample (black hatched) than in control ($f_X'$; red), out to $d$\,$\approx$\,2\,kpc. The uncertainties show Poisson sampling errors on the primary sample.}
\label{fig:xf}
\end{figure}

The \chandra\ cross-match resulted in over 6,500 X-ray detected sources in the primary sample and about 7,400 in control. Importantly, this translates into a very significant difference in terms of population fractions. The fraction of X-ray detected sources ($f_X$)\,=\,3.52\,($\pm$\,0.04)\,\pertenmille{} %2,763 (4.3\,\pertenmille{})
for the primary sample.\footnote{Unless otherwise stated, statistical uncertainties on population fractions throughout the paper refer to 68\%\ confidence Poisson limits, and are appropriate for small number statistics \citep{gehrels86}.} This is a factor of $\approx$\,4.5\,times higher than that in control, $f_X'$\,=\,0.77\,($\pm$\,0.01\,\pertenmille{}), a difference that holds true across all mags and much of distance range probed. This is illustrated in Fig.\,\ref{fig:xf}, showing $f_X$ split as a function of distance, with $f_X$ being consistently higher than $f_X'$ out to about 2 kpc, reflecting the drop in the joint X-ray detection and astrometric selection efficiencies with distance. The full catalogues are available through CDS\footnote{Reference link to be added upon publication}.

\begin{figure}
\hspace*{-0.7cm}
	\includegraphics[angle=90,width=1.1\columnwidth]{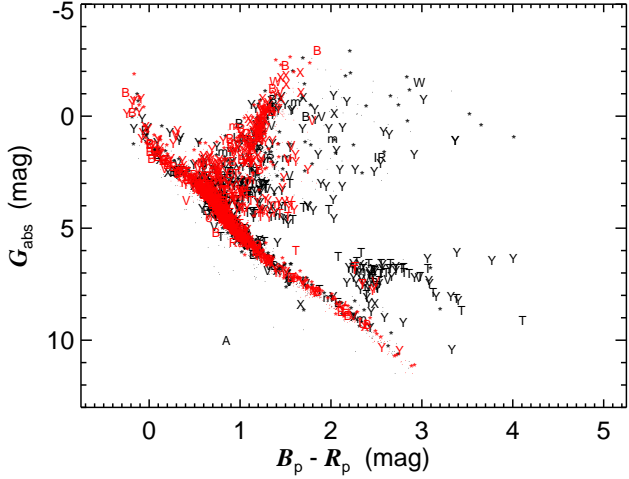}
	\caption{\gaia\ colour--mag diagram for the X-ray cross-matched samples (black: primary sample; red: control sample). Individual object types from {\tt SIMBAD} are denoted, with a few key types as follows. `B': Binary, `V': Variable, `Y': YSO, `T': TTau, `X': X-ray source, `W': Wolf-Rayet, `m': Emission line source. Objects with a more normal stellar classification (e.g. `Star') are denoted by an asterisk, and those without an archival class are denoted as dots.}
\label{fig:hr}
\end{figure}

Fig.\,\ref{fig:hr} shows the \gaia\ colour--mag diagram (CMD) for the X-ray cross-matched samples. EDR3 does not report extinction and reddening values, so we extracted and applied these from DR2, where available. We advise caution that these saturate around $A_{\rm G}$\,$\approx$\,3\,mag and $E(B_{\rm p}-R_{\rm p})$\,$\approx$\,1.5\,mag \citep{andrae18};  thus, these corrections are underestimated for many individual systems. Nevertheless, canonical features such as the main sequence (MS) and the giant branch immediately stand out. In addition, a clump redder than the MS at relatively faint levels is apparent, corresponding to the expected locus of young stellar objects (YSOs), or to the less-understood population of sub-subgiants \cite[e.g. ][]{geller17}.

There are some important differences apparent between the two samples. The primary sample extends substantially deeper into the evolved/reddened branch, off the MS, than the control sample. Specifically, the mean $\langle$\bprp$\rangle$ colour is $\approx$\,0.3\,mag redder in primary than control. The reddest source has a \bprp\ colour almost 2\,mag larger than in control. All objects plotted in the CMD have reddening and extinction corrections applied, so underestimates in these corrections likely only play a partial role in explaining these differences. Instead, the larger scatter of primary sample sources suggests that \noise-selection preferentially picks up objects with a wide spread of evolutionary phases and source classes. 

Source type assignments are denoted on the diagram, where available. Information regarding known object classifications was collated using a simple cross-match with {\tt SIMBAD}\footnote{\url{http://simbad.u-strasbg.fr/simbad/}}. The `{\tt main\_type}' of the closest association within a threshold distance of 1\,arcsec of the \gaia\ coordinates was extracted. These were then grouped into a few broad categories for summarisation. The complete list of classes and categories can be found in the Appendix. Sky coverage of classifications is patchy and highly incomplete. But this exercise is solely meant to provide first insight into the putative nature of our selected population. 

\begin{figure}
\hspace*{-0.25cm}
	\includegraphics[angle=0,width=1.1\columnwidth]{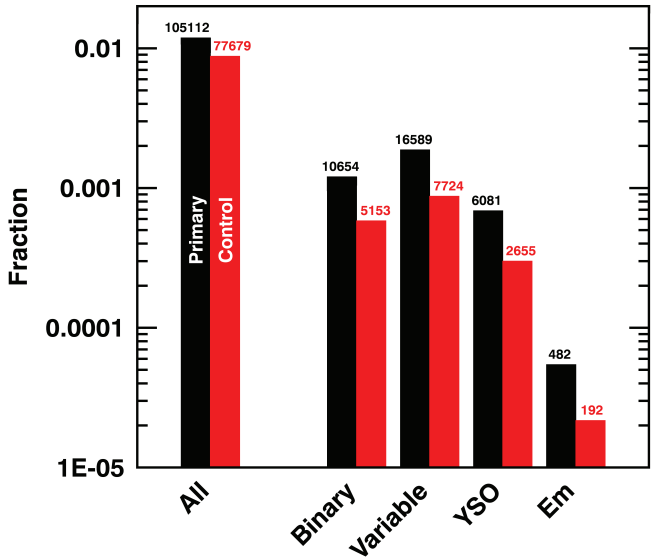}
	\caption{Distribution of object classes cross-matched with {\tt SIMBAD}. Here, distance-matching has been applied for a fair cross-comparison between primary and control. Only a few broad categories are shown (for the full list, we refer the reader to the Appendix). Above each histogram bar, the number of objects in that bin is stated.}
\label{fig:barplot}
\end{figure}

About 1\%\ of objects from the full \gaia-only selection have a documented classification in {\tt SIMBAD},\footnote{as of September, 2021.} albeit being in uncertain in many cases. Since object type determination may be implicitly distance or mag dependent, a fair comparison was carried out by distance-matching the two samples. For this, we randomly selected one control sample source for every primary sample source, to within a distance threshold of 0.05\,kpc, so that the distribution of distances becomes statistically indistinguishable. The resultant distance-matched sub-sample comprised about 8 million objects, and results on the relative comparison are shown in Fig.\,\ref{fig:barplot} and Table\,\ref{tab:fractions}. 

The fraction of objects with known classifications is $\approx$\,0.9\% (control) and 1.2\% (primary). By contrast, the fraction of known binaries in the primary sample outnumbers those in control by more than a factor of 2. These include non-interacting systems as well as interacting binaries such as XRBs and CVs. Three other source types are highlighted here -- variables, emission line objects, and young stellar objects -- as these will be relevant to the Discussion later. In all these cases, again, the corresponding fraction of systems in the primary sample is more than a factor of 2 larger than control. 

In X-rays, the matched sample luminosity distributions are qualitatively similar, peaking close to $\langle {\rm log}[L_{\rm X}\,/\,{\rm erg}\,{\rm s}^{-1}] \rangle$\,=\,29.5, with a high tail extending beyond a peak of $L_{\rm X}$\,$\approx$\,10$^{32}$\,erg\,s$^{-1}$.

\begin{table}
    \centering
    {\tt SIMBAD} object type distributions
    \begin{tabular}{lcc}
\hline
\hline
Class & $f_{\rm Primary}$ & $f^{'}_{\rm Control}$\\
      &  10$^{-4}$        & 10$^{-4}$\\
      \hline
    All& 118.7\,$\pm$\,0.4  & 87.8\,$\pm$\,0.3 \\  
    &&\\
    Binary (B)& 12.0\,$\pm$\,0.1    & 5.8\,$\pm$\,0.1\\
    Variable (V)& 18.7\,$\pm$\,0.1 & 8.7\,$\pm$\,0.1 \\
    YSO (Y)& 6.9\,$\pm$\,0.1    & 3.0\,$\pm$\,0.1\\
    Emission line (m)& 0.54\,$\pm$\,0.02    & 0.22\,$\pm$\,0.02\\
    \hline
    \end{tabular}
    \caption{Fractions of object types from {\tt SIMBAD}, split into a few of the key broad categories. These refer to the distance-matched samples for fair comparison, and shown pictorially in Fig.\,\ref{fig:barplot}. The letter in parentheses is that used in the CMD (Fig.\,\ref{fig:hr}) to depict them.     \label{tab:fractions}}
\end{table}

\section{Discussion}

We began this study with the aim of searching for previously unidentified or ill characterised, putative interacting binaries. What is the nature of our sample, and are we likely to have detected robust accreting binary candidates?

Our prime selection criterion is {\em astrometry}-based (choosing a sample of Galactic point sources with significant astrometric excess noise; \noise), which was then refined using {\em photometric} properties (X-ray detection). The nature of \noise\ selection and its veracity still remain unclear in the early \gaia\ data releases, and the individual astrometric measurements have not yet been released. This is why we chose to explore the influence of \noise\ selection in a controlled manner and in a relative sense between two closely similar samples. The most significant highlights of our selection are: 

\begin{enumerate}
    
    \item The X-ray detection fraction with \chandra\ in our primary sample is significantly higher, about 5 times more, than control (Fig.\,\ref{fig:xf}, Table\,\ref{tab:stats}). This is true across all (optical) mags and out to distances of $\approx$\,2\,kpc where the samples become comparable presumably due to the currently limited X-ray sensitivity and a decreasing \noise\ measurement efficacy with distance. 
    
    \item Sources with significant \noise\ lie closer to us than the control sample ($\langle d'\rangle$\,=\,1.9\,kpc vs. $\langle d\rangle$\,=\,1.2\,kpc), again, presumably due to the fact that higher order astrometric signatures are easier to measure for nearer (brighter) objects.

    \item Occupancy in colour--mag space across the CMD differs between the two samples, with the primary sample occupying the regime of redder/more evolved colours (Fig.\,\ref{fig:hr}). 

    \item We find a higher fraction of objects with current classified types as binaries, variables, emission line sources and young stellar objects in the primary sample (Fig.\,\ref{fig:barplot} and Table\,\ref{tab:fractions}).

\end{enumerate}

\noindent
While any one of the above differences may be attributable to selection effects or systematic uncertainties in astrometric fitting, such explanations do not suffice when considering the above differences cumulatively. In particular, systematic effects in any one mission or wavelength (e.g. \gaia) would not obviously be expected to result in differences at other wavelengths (\chandra) or in other catalogues ({\tt SIMBAD} spectroscopic classifications). Thus, the above differences suggest that \noise\ selection is  effective in picking up sources with intrinsically distinct properties, on average. 

Our X-ray selection is able to probe down to the level where quiescent emission from XRBs is detectable ($L_{\rm X}$\,$\gtsim$\,10$^{28-32}$\,erg\,s$^{-1}$). Accreting XRBs often comprise evolved donor stars undergoing Roche-lobe-overflow mass transfer that generates X-rays, also encouragingly consistent with the larger fraction of objects located off the main sequence in the CMD. All these facts are consistent with the presence of new quiescent accreting binaries amongst our primary sample. But this is not unambiguous. 

Regarding known source classifications, {\em emission line objects} with limited prior observational follow-up could also be hiding accreting systems with viscous disc heating or with irradiation powering line emission. Variability is an additional characteristic property of quiescent accretion (e.g. \citealt{zurita03}). {\em Variables}, however, cover a broad range of source classes, and we cannot rule out the possibility that these may include single active stars where the variability causes systematic perturbations to the individual astrometric measurements, resulting in artificially boosted \noise. The presence of {\em YSOs} amongst our selected objects may appear surprising at first. But this could be a simple consequence of the fact that YSOs have a high multiplicity that decreases with evolutionary phase and scales with mass \citep[e.g. ][]{pomohaci19}. Consequently, younger YSOs in our primary sample could well include a high fraction of binaries showing significant \noise. YSOs are known to be variable in flux, and are also known to be X-ray sources \citep{feigelson99}. Thus, there are multiple reasons why our selection picks up YSOs.

A final systematic issue to be aware of is that of crowding. While we have attempted to mitigate this issue by excluding close \gaia\ neighbours, it is not inconceivable that other contaminants (those below the \gaia\ pipeline detection threshold) may be impacting source astrometrics in regions such as star and globular clusters. Partially resolved double stars could also introduce biases in single-star astrometric solutions. \citet{fabricius21} suggest cuts on the {\tt Image Parameter Determination (ipd)} factors {\tt ipd\_frac\_multi\_peak\,$>$2 OR ipd\_gof\_harmonic\_amplitude\,$>$\,0.1}, in order to flag solutions with multiple peaks and asymmetries. Using these criteria would flag 22\,\% of our primary sample as being potentially impacted due to the presence of partially resolved doubles. While we have not included blanket additional cuts to remove such objects (since they could include objects of inherent interest for our selection), any inference regarding the nature of objects in particularly crowded regions should be treated carefully. The bulk of the sample is not impacted by such {\tt ipd} flags.

Radial velocity curves would be needed to confirm the nature of these various object types, to test for binarity and to measure their system characteristics, while deeper X-ray and radio data could establish the nature of high-energy activity. Current model predictions suggest that \gaia\ ought to detect several hundreds to thousands of BHs in binary systems \citep[e.g. ][]{yamaguchi18, breivik17, chawla21}, with a preference for more precise measurements of longer period systems \citep{andrews19}. Methods have also been proposed to detect non-interacting systems with MS companions \citep{shahaf19} with \gaia. Quiescently accreting systems (of the kind that we have discussed herein) will likely be a fraction of these, but will be the `low hanging fruit' that are likely to be the easiest ones to find, follow up and confirm. 

\begin{figure}
\hspace*{-0.7cm}
	\includegraphics[angle=90,width=1.1\columnwidth]{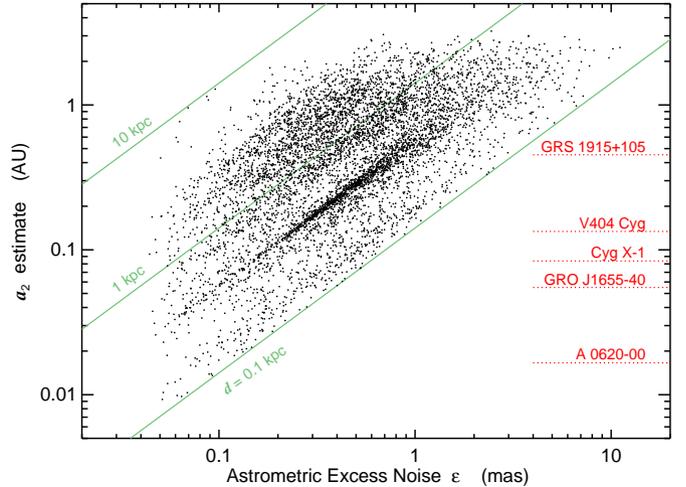}
	\caption{Astrometric excess noise (\noise) vs. putative binary companion star semi-major axis $a_2$. The latter is computed from Eq.\,\ref{eq:wobble} and assuming that \noise\ is an estimator of \w/$\sqrt{2}$ in the limit of perfect sampling. See text for details and caveats. Three loci covering a range of distances are plotted in green. Known or estimated values of $a_2$ for five known BHXBs are denoted by the red dotted lines. The clustering of objects near the middle of the figure is likely an artefact of dense X-ray sampling of the Orion star-forming region at $d$\,$\approx$\,0.4\,kpc.
\label{fig:a2}}
\end{figure}

The mean value of excess noise across our primary X-ray detected sample, $\langle \epsilon \rangle$\,=\,0.44$_{-0.26}^{+0.64}$\,mas at the mean distance of 0.9 kpc, translates to an expected orbital semi-major axis $a_2$\,$\approx$\,0.5\,AU if interpreted as the maximal astrometric orbital perturbation of a binary projected onto the sky. Any binary systems with these characteristic sizes would be akin to long period systems such as the long-period BHXB GRS\,1915+105 \citep{casaresjonker14} or the cataclysmic variable T\,CrB \citep{fekel00}. The most compact systems amongst our sample, however, extend down to $a_2$\,$\approx$\,0.009\,AU, easily compatible with the regime of short-period accreting binaries \citep[e.g. ][]{casaresjonker14}. This is illustrated in Fig.\,\ref{fig:a2}, where the estimated values of $a_2$ are plotted for our full primary sample of $\approx$\,6,500 objects as a function of \noise, and compared with $a_2$ estimates for several known BHXBs which cover similar parameter space.

These are simple first-order $a_2$ estimates, and the aforementioned caveats (assumption of a circular orbit, well-sampled astrometry, and no radiative contribution from the accreting primary component) should be kept in mind while drawing any detailed inferences on individual systems. Additional caution is also warranted because the current magnitudes of \noise\ actually {\em overestimate} expectations in a few known systems. 
Examples include Her\,X--1 (\noise\,=\,0.09\,\p\,0.01\,mas as compared to $\hat{\omega}$\,=\,0.002\,mas) and V404\,Cyg (\noise\,=\,0.41\,\p\,0.08\,mas as opposed to $\hat{\omega}$\,=\,0.04\,mas).\footnote{All physical parameters for known systems quoted here and in Fig.\,\ref{fig:a2} are from previous works including \citet{reynolds97}, \citet{casaresjonker14}, and updated with recent  distances \citep{reid14, gandhi19, millerjones21}.} 
One possible systematic issue here may be photometric variability. V404\,Cyg underwent a dramatic outburst in 2015, displaying prolific flux changes by up to 7\,mag over a period of a few weeks \citep[e.g., ][]{kimura16, g16_v404}. But given that the outburst of V404\,Cyg was relatively brief, it is difficult to see how such changes could dominate the astrometric solution determined over the full EDR3 observation period. Her\,X--1 is an eclipsing system with known orbital and superorbital flux modulations on characteristic timescales of about 2\,days and 1\,month, respectively \citep{jurua11}. Such variations could potentially introduce systematic variations in the epochwise astrometric uncertainties. 
A second systematic issue may come down to an {\em underestimate} of the pipeline parallax uncertainties, as suggested by \cite{elbadry21_binaries}; this would artificially boost \noise, though the root cause of such an underestimate of the parallax uncertainties remains unclear. Finally, attitude errors could also bias the astrometric solutions; attitude errors are time-dependent and have been globally adjusted for weighted residuals in released pipeline solutions \citep{lindegren12, lindegren18, lindegren21}. The individual astrometric measurements, not yet available, will be needed to clarify the underlying cause of the mismatch between expectations and the pipeline reported measurements.

\section{Conclusions}

We have explored the statistics and nature of objects found using astrometric excess noise selection in \gaia\ EDR3. Our initial aim was to identify candidate accreting systems. But a variety of tests carried out in a controlled fashion demonstrate that excess noise selection is effective in identifying a diverse range of active source classes. X-ray cross-matching is used to refine the selection to identify putative quiescent interacting binaries, variables, emission line sources and young stellar objects, amongst others. 

Our astrometric selection encompasses the parameter space expected for orbital wobble in accreting binaries. But  caution is needed when interpreting the current reported values of excess noise, especially when \noise\ is small (well below 1\,mas); systematic effects including attitude errors, partially resolved double stars, and source variability likely contaminate our selection to a certain extent at present. Full astrometric solutions in future data releases will help to understand the selection function more quantitatively. Nevertheless, our sample is unlikely to be dominated by such systematics, because of the addition of independent X-ray constraints. Upcoming all-sky X-ray followup from \erosita\ will also provide a treasure trove of other candidate active systems enhancing the sample that we present here \citep{erosita}. Similarly, in the future, Galactic plane follow-up with the ngVLA should accomplish the same in the radio \citep{maccaronengvla}. 

This is a first detailed attempt to utilise astrometric wobble to search for robust binary candidates over the full sky. Our sample is not meant to be complete in any physical sense yet; rather, this is a first attempt to pick the `low hanging' fruit in EDR3 astrometric noise selection of accreting binaries. We release our primary source sample to enable multiwavelength follow-up and characterisation of individual systems. 

\section*{Acknowledgements}

PG, JAP and YZ acknowledge support from STFC and a UGC-UKIERI Thematic Partnership. DAHB acknowledges research support from the South African National Research Foundation. PG is indebted to the {\em eROSITA} team, in particular J. Wilms and A. Schwope, for collaboration on a related expanded project. He also thanks J. Tomsick, W.-M. Gu, and T. Yi for discussions. We are grateful to the anonymous reviewers for helping to provide substantial focus to improve the presentation of our results.

This work has made use of data from the European Space Agency (ESA) mission Gaia (https://www.cosmos.esa.int/gaia), processed by the Gaia Data Processing and Analysis Consortium (DPAC, https://www.cosmos.esa.int/web/gaia/dpac/consortium). Funding for the DPAC has been provided by national institutions, in particular the institutions participating in the Gaia Multilateral Agreement.

\chandra\ is a mission of the National Aeronautics and Space Agency (NASA). This research has made use of data obtained from the Chandra Source Catalog, provided by the Chandra X-ray Center (CXC) as part of the Chandra Data Archive.

We are grateful to both the \gaia\ and the \chandra\ helpdesk teams for patiently answering our questions on database queries. PG also acknowledges F.\,Arenou for helpful discussions regarding the impact of partly resolved double stars on \noise\ in \gaia\ EDR3.

Extensive use was made of the {\tt TOPCAT} software \citep{topcat}, and the IDL {\tt astrolib} routines herein  \citep{idlastrolib}. This research has made use of the {\tt SIMBAD} database,
operated at CDS, Strasbourg, France, `The SIMBAD astronomical database' \citep{simbad}.

This work was carried out during the COVID-19 pandemic and we are indebted to all those working to keep the public safe during these times. 

\section*{Data availability}

All of the multiwavelength catalogues utilised herein (\gaia,  \chandra) are publicly available. Our final extracted catalogue will also be available through Vizier/CDS upon publication of this manuscript.



\bibliographystyle{mnras}
\bibliography{gandhi22} % if your bibtex file is called example.bib

\begin{thebibliography}{}
\makeatletter
\relax
\def\mn@urlcharsother{\let\do\@makeother \do\$\do\&\do\#\do\^\do\_\do\%\do\~}
\def\mn@doi{\begingroup\mn@urlcharsother \@ifnextchar [ {\mn@doi@}
  {\mn@doi@[]}}
\def\mn@doi@[#1]#2{\def\@tempa{#1}\ifx\@tempa\@empty \href
  {http://dx.doi.org/#2} {doi:#2}\else \href {http://dx.doi.org/#2} {#1}\fi
  \endgroup}
\def\mn@eprint#1#2{\mn@eprint@#1:#2::\@nil}
\def\mn@eprint@arXiv#1{\href {http://arxiv.org/abs/#1} {{\tt arXiv:#1}}}
\def\mn@eprint@dblp#1{\href {http://dblp.uni-trier.de/rec/bibtex/#1.xml}
  {dblp:#1}}
\def\mn@eprint@#1:#2:#3:#4\@nil{\def\@tempa {#1}\def\@tempb {#2}\def\@tempc
  {#3}\ifx \@tempc \@empty \let \@tempc \@tempb \let \@tempb \@tempa \fi \ifx
  \@tempb \@empty \def\@tempb {arXiv}\fi \@ifundefined
  {mn@eprint@\@tempb}{\@tempb:\@tempc}{\expandafter \expandafter \csname
  mn@eprint@\@tempb\endcsname \expandafter{\@tempc}}}

\bibitem[\protect\citeauthoryear{{Abbott} et~al.,}{{Abbott}
  et~al.}{2016}]{ligo1}
{Abbott} B.~P.,  et~al., 2016, \mn@doi [\prl] {10.1103/PhysRevLett.116.061102},
  \href {https://ui.adsabs.harvard.edu/abs/2016PhRvL.116f1102A} {116, 061102}

\bibitem[\protect\citeauthoryear{{Abdul-Masih} et~al.,}{{Abdul-Masih}
  et~al.}{2020}]{Abdul-Masih20}
{Abdul-Masih} M.,  et~al., 2020, \mn@doi [\nat] {10.1038/s41586-020-2216-x},
  \href {https://ui.adsabs.harvard.edu/abs/2020Natur.580E..11A} {580, E11}

\bibitem[\protect\citeauthoryear{{Andrae} et~al.,}{{Andrae}
  et~al.}{2018}]{andrae18}
{Andrae} R.,  et~al., 2018, \mn@doi [\aap] {10.1051/0004-6361/201732516}, \href
  {https://ui.adsabs.harvard.edu/abs/2018A&A...616A...8A} {616, A8}

\bibitem[\protect\citeauthoryear{{Andrews}, {Breivik}  \&
  {Chatterjee}}{{Andrews} et~al.}{2019}]{andrews19}
{Andrews} J.~J.,  {Breivik} K.,   {Chatterjee} S.,  2019, \mn@doi [\apj]
  {10.3847/1538-4357/ab441f}, \href
  {https://ui.adsabs.harvard.edu/abs/2019ApJ...886...68A} {886, 68}

\bibitem[\protect\citeauthoryear{{Barstow} et~al.,}{{Barstow}
  et~al.}{2014}]{barstow14_gaia_wp}
{Barstow} M.~A.,  et~al., 2014, arXiv e-prints, \href
  {https://ui.adsabs.harvard.edu/abs/2014arXiv1407.6163B} {p. arXiv:1407.6163}

\bibitem[\protect\citeauthoryear{{Belokurov} et~al.,}{{Belokurov}
  et~al.}{2020}]{belokurov20}
{Belokurov} V.,  et~al., 2020, \mn@doi [\mnras] {10.1093/mnras/staa1522}, \href
  {https://ui.adsabs.harvard.edu/abs/2020MNRAS.496.1922B} {496, 1922}

\bibitem[\protect\citeauthoryear{{Bodensteiner} et~al.,}{{Bodensteiner}
  et~al.}{2020}]{bodensteiner20}
{Bodensteiner} J.,  et~al., 2020, \mn@doi [\aap] {10.1051/0004-6361/202038682},
  \href {https://ui.adsabs.harvard.edu/abs/2020A&A...641A..43B} {641, A43}

\bibitem[\protect\citeauthoryear{{Breivik}, {Chatterjee}  \&
  {Larson}}{{Breivik} et~al.}{2017}]{breivik17}
{Breivik} K.,  {Chatterjee} S.,   {Larson} S.~L.,  2017, \mn@doi [\apjl]
  {10.3847/2041-8213/aa97d5}, \href
  {https://ui.adsabs.harvard.edu/abs/2017ApJ...850L..13B} {850, L13}

\bibitem[\protect\citeauthoryear{{Casares} \& {Jonker}}{{Casares} \&
  {Jonker}}{2014}]{casaresjonker14}
{Casares} J.,  {Jonker} P.~G.,  2014, \mn@doi [\ssr]
  {10.1007/s11214-013-0030-6}, \href
  {https://ui.adsabs.harvard.edu/abs/2014SSRv..183..223C} {183, 223}

\bibitem[\protect\citeauthoryear{{Chawla}, {Chatterjee}, {Breivik}, {Krishna
  Moorthy}, {Andrews}  \& {Sanderson}}{{Chawla} et~al.}{2021}]{chawla21}
{Chawla} C.,  {Chatterjee} S.,  {Breivik} K.,  {Krishna Moorthy} C.,  {Andrews}
  J.~J.,   {Sanderson} R.~E.,  2021, arXiv e-prints, \href
  {https://ui.adsabs.harvard.edu/abs/2021arXiv211005979C} {p. arXiv:2110.05979}

\bibitem[\protect\citeauthoryear{{Corral-Santana}, {Casares},
  {Mu{\~n}oz-Darias}, {Bauer}, {Mart{\'\i}nez-Pais}  \&
  {Russell}}{{Corral-Santana} et~al.}{2016}]{blackcat}
{Corral-Santana} J.~M.,  {Casares} J.,  {Mu{\~n}oz-Darias} T.,  {Bauer} F.~E.,
  {Mart{\'\i}nez-Pais} I.~G.,   {Russell} D.~M.,  2016, \mn@doi [\aap]
  {10.1051/0004-6361/201527130}, \href
  {https://ui.adsabs.harvard.edu/abs/2016A&A...587A..61C} {587, A61}

\bibitem[\protect\citeauthoryear{{El-Badry} \& {Burdge}}{{El-Badry} \&
  {Burdge}}{2021}]{elbadryburdge21}
{El-Badry} K.,  {Burdge} K.,  2021, arXiv e-prints, \href
  {https://ui.adsabs.harvard.edu/abs/2021arXiv211107925E} {p. arXiv:2111.07925}

\bibitem[\protect\citeauthoryear{{El-Badry} \& {Quataert}}{{El-Badry} \&
  {Quataert}}{2021}]{elbadry21_hr6819}
{El-Badry} K.,  {Quataert} E.,  2021, \mn@doi [\mnras] {10.1093/mnras/stab285},
  \href {https://ui.adsabs.harvard.edu/abs/2021MNRAS.502.3436E} {502, 3436}

\bibitem[\protect\citeauthoryear{{El-Badry}, {Rix}  \& {Heintz}}{{El-Badry}
  et~al.}{2021}]{elbadry21_binaries}
{El-Badry} K.,  {Rix} H.-W.,   {Heintz} T.~M.,  2021, \mn@doi [\mnras]
  {10.1093/mnras/stab323}, \href
  {https://ui.adsabs.harvard.edu/abs/2021MNRAS.506.2269E} {506, 2269}

\bibitem[\protect\citeauthoryear{{Eldridge}, {Stanway}, {Breivik}, {Casey},
  {Steeghs}  \& {Stevance}}{{Eldridge} et~al.}{2020}]{eldridge20}
{Eldridge} J.~J.,  {Stanway} E.~R.,  {Breivik} K.,  {Casey} A.~R.,  {Steeghs}
  D.~T.~H.,   {Stevance} H.~F.,  2020, \mn@doi [\mnras]
  {10.1093/mnras/staa1324}, \href
  {https://ui.adsabs.harvard.edu/abs/2020MNRAS.495.2786E} {495, 2786}

\bibitem[\protect\citeauthoryear{{Evans} et~al.,}{{Evans} et~al.}{2010}]{csc}
{Evans} I.~N.,  et~al., 2010, \mn@doi [\apjs] {10.1088/0067-0049/189/1/37},
  \href {https://ui.adsabs.harvard.edu/abs/2010ApJS..189...37E} {189, 37}

\bibitem[\protect\citeauthoryear{{Evans} et~al.,}{{Evans} et~al.}{2020}]{csc2}
{Evans} I.~N.,  et~al., 2020, in American Astronomical Society Meeting
  Abstracts. American Astronomical Society Meeting Abstracts.
p. 154.05

\bibitem[\protect\citeauthoryear{{Fabricius} et~al.,}{{Fabricius}
  et~al.}{2021}]{fabricius21}
{Fabricius} C.,  et~al., 2021, \mn@doi [\aap] {10.1051/0004-6361/202039834},
  \href {https://ui.adsabs.harvard.edu/abs/2021A&A...649A...5F} {649, A5}

\bibitem[\protect\citeauthoryear{{Feigelson} \& {Montmerle}}{{Feigelson} \&
  {Montmerle}}{1999}]{feigelson99}
{Feigelson} E.~D.,  {Montmerle} T.,  1999, \mn@doi [\araa]
  {10.1146/annurev.astro.37.1.363}, \href
  {https://ui.adsabs.harvard.edu/abs/1999ARA&A..37..363F} {37, 363}

\bibitem[\protect\citeauthoryear{{Fekel}, {Joyce}, {Hinkle}  \&
  {Skrutskie}}{{Fekel} et~al.}{2000}]{fekel00}
{Fekel} F.~C.,  {Joyce} R.~R.,  {Hinkle} K.~H.,   {Skrutskie} M.~F.,  2000,
  \mn@doi [\aj] {10.1086/301260}, \href
  {https://ui.adsabs.harvard.edu/abs/2000AJ....119.1375F} {119, 1375}

\bibitem[\protect\citeauthoryear{{Fishbach} \& {Kalogera}}{{Fishbach} \&
  {Kalogera}}{2021}]{fishbach21}
{Fishbach} M.,  {Kalogera} V.,  2021, arXiv e-prints, \href
  {https://ui.adsabs.harvard.edu/abs/2021arXiv211102935F} {p. arXiv:2111.02935}

\bibitem[\protect\citeauthoryear{{Gaia Collaboration} et~al.,}{{Gaia
  Collaboration} et~al.}{2018}]{gaiadr2}
{Gaia Collaboration} et~al., 2018, \mn@doi [\aap]
  {10.1051/0004-6361/201833051}, \href
  {http://adsabs.harvard.edu/abs/2018A%26A...616A...1G} {616, A1}

\bibitem[\protect\citeauthoryear{{Gaia Collaboration} et~al.,}{{Gaia
  Collaboration} et~al.}{2021}]{gaiaedr3}
{Gaia Collaboration} et~al., 2021, \mn@doi [\aap]
  {10.1051/0004-6361/202039657}, \href
  {https://ui.adsabs.harvard.edu/abs/2021A&A...649A...1G} {649, A1}

\bibitem[\protect\citeauthoryear{{Gandhi} et~al.,}{{Gandhi}
  et~al.}{2016}]{g16_v404}
{Gandhi} P.,  et~al., 2016, \mn@doi [\mnras] {10.1093/mnras/stw571}, \href
  {http://adsabs.harvard.edu/abs/2016MNRAS.459..554G} {459, 554}

\bibitem[\protect\citeauthoryear{{Gandhi}, {Rao}, {Johnson}, {Paice}  \&
  {Maccarone}}{{Gandhi} et~al.}{2019}]{gandhi19}
{Gandhi} P.,  {Rao} A.,  {Johnson} M.~A.~C.,  {Paice} J.~A.,   {Maccarone}
  T.~J.,  2019, \mn@doi [\mnras] {10.1093/mnras/stz438}, \href
  {http://adsabs.harvard.edu/abs/2019MNRAS.485.2642G} {485, 2642}

\bibitem[\protect\citeauthoryear{{Gandhi}, {Rao}, {Charles}, {Belczynski},
  {Maccarone}, {Arur}  \& {Corral-Santana}}{{Gandhi} et~al.}{2020}]{gandhi20}
{Gandhi} P.,  {Rao} A.,  {Charles} P.~A.,  {Belczynski} K.,  {Maccarone} T.~J.,
   {Arur} K.,   {Corral-Santana} J.~M.,  2020, \mn@doi [\mnras]
  {10.1093/mnrasl/slaa081}, \href
  {https://ui.adsabs.harvard.edu/abs/2020MNRAS.496L..22G} {496, L22}

\bibitem[\protect\citeauthoryear{{Gehrels}}{{Gehrels}}{1986}]{gehrels86}
{Gehrels} N.,  1986, \mn@doi [\apj] {10.1086/164079}, \href
  {https://ui.adsabs.harvard.edu/abs/1986ApJ...303..336G} {303, 336}

\bibitem[\protect\citeauthoryear{{Geller} et~al.,}{{Geller}
  et~al.}{2017}]{geller17}
{Geller} A.~M.,  et~al., 2017, \mn@doi [\apj] {10.3847/1538-4357/aa6af3}, \href
  {https://ui.adsabs.harvard.edu/abs/2017ApJ...840...66G} {840, 66}

\bibitem[\protect\citeauthoryear{{Gould} \& {Salim}}{{Gould} \&
  {Salim}}{2002}]{gould02}
{Gould} A.,  {Salim} S.,  2002, \mn@doi [\apj] {10.1086/340435}, \href
  {https://ui.adsabs.harvard.edu/abs/2002ApJ...572..944G} {572, 944}

\bibitem[\protect\citeauthoryear{{G{\"u}del}}{{G{\"u}del}}{2004}]{Gudel04}
{G{\"u}del} M.,  2004, \mn@doi [\aapr] {10.1007/s00159-004-0023-2}, \href
  {https://ui.adsabs.harvard.edu/abs/2004A&ARv..12...71G} {12, 71}

\bibitem[\protect\citeauthoryear{{Irrgang}, {Geier}, {Kreuzer}, {Pelisoli}  \&
  {Heber}}{{Irrgang} et~al.}{2020}]{irrgang20}
{Irrgang} A.,  {Geier} S.,  {Kreuzer} S.,  {Pelisoli} I.,   {Heber} U.,  2020,
  \mn@doi [\aap] {10.1051/0004-6361/201937343}, \href
  {https://ui.adsabs.harvard.edu/abs/2020A&A...633L...5I} {633, L5}

\bibitem[\protect\citeauthoryear{{Jayasinghe} et~al.,}{{Jayasinghe}
  et~al.}{2021}]{Jayasinghe21}
{Jayasinghe} T.,  et~al., 2021, \mn@doi [\mnras] {10.1093/mnras/stab907}, \href
  {https://ui.adsabs.harvard.edu/abs/2021MNRAS.504.2577J} {504, 2577}

\bibitem[\protect\citeauthoryear{{Johnson}, {Gandhi}, {Chapman}, {Moreau},
  {Charles}, {Clarkson}  \& {Hill}}{{Johnson} et~al.}{2019}]{johnson19}
{Johnson} M. A.~C.,  {Gandhi} P.,  {Chapman} A.~P.,  {Moreau} L.,  {Charles}
  P.~A.,  {Clarkson} W.~I.,   {Hill} A.~B.,  2019, \mn@doi [\mnras]
  {10.1093/mnras/sty3466}, \href
  {https://ui.adsabs.harvard.edu/abs/2019MNRAS.484...19J} {484, 19}

\bibitem[\protect\citeauthoryear{{Jonker}, {Kaur}, {Stone}  \&
  {Torres}}{{Jonker} et~al.}{2021}]{jonker21}
{Jonker} P.~G.,  {Kaur} K.,  {Stone} N.,   {Torres} M. A.~P.,  2021, \mn@doi
  [\apj] {10.3847/1538-4357/ac2839}, \href
  {https://ui.adsabs.harvard.edu/abs/2021ApJ...921..131J} {921, 131}

\bibitem[\protect\citeauthoryear{{Jurua}, {Charles}, {Still}  \&
  {Meintjes}}{{Jurua} et~al.}{2011}]{jurua11}
{Jurua} E.,  {Charles} P.~A.,  {Still} M.,   {Meintjes} P.~J.,  2011, \mn@doi
  [\mnras] {10.1111/j.1365-2966.2011.19494.x}, \href
  {https://ui.adsabs.harvard.edu/abs/2011MNRAS.418..437J} {418, 437}

\bibitem[\protect\citeauthoryear{{Kimura} et~al.,}{{Kimura}
  et~al.}{2016}]{kimura16}
{Kimura} M.,  et~al., 2016, \mn@doi [\nat] {10.1038/nature16452}, \href
  {http://adsabs.harvard.edu/abs/2016Natur.529...54K} {529, 54}

\bibitem[\protect\citeauthoryear{{Landsman}}{{Landsman}}{1993}]{idlastrolib}
{Landsman} W.~B.,  1993, in {Hanisch} R.~J.,  {Brissenden} R.~J.~V.,   {Barnes}
  J.,  eds,  Astronomical Society of the Pacific Conference Series Vol. 52,
  Astronomical Data Analysis Software and Systems II. p.~246

\bibitem[\protect\citeauthoryear{{Lindegren}, {Lammers}, {Hobbs}, {O'Mullane},
  {Bastian}  \& {Hern{\'a}ndez}}{{Lindegren} et~al.}{2012}]{lindegren12}
{Lindegren} L.,  {Lammers} U.,  {Hobbs} D.,  {O'Mullane} W.,  {Bastian} U.,
  {Hern{\'a}ndez} J.,  2012, \mn@doi [\aap] {10.1051/0004-6361/201117905},
  \href {https://ui.adsabs.harvard.edu/abs/2012A&A...538A..78L} {538, A78}

\bibitem[\protect\citeauthoryear{{Lindegren} et~al.,}{{Lindegren}
  et~al.}{2018}]{lindegren18}
{Lindegren} L.,  et~al., 2018, \mn@doi [\aap] {10.1051/0004-6361/201832727},
  \href {https://ui.adsabs.harvard.edu/abs/2018A&A...616A...2L} {616, A2}

\bibitem[\protect\citeauthoryear{{Lindegren} et~al.,}{{Lindegren}
  et~al.}{2021a}]{lindegren21}
{Lindegren} L.,  et~al., 2021a, \mn@doi [\aap] {10.1051/0004-6361/202039709},
  \href {https://ui.adsabs.harvard.edu/abs/2021A&A...649A...2L} {649, A2}

\bibitem[\protect\citeauthoryear{{Lindegren} et~al.,}{{Lindegren}
  et~al.}{2021b}]{lindegren21_zpoffset}
{Lindegren} L.,  et~al., 2021b, \mn@doi [\aap] {10.1051/0004-6361/202039653},
  \href {https://ui.adsabs.harvard.edu/abs/2021A&A...649A...4L} {649, A4}

\bibitem[\protect\citeauthoryear{{Liu} et~al.,}{{Liu} et~al.}{2019}]{liu19_lb1}
{Liu} J.,  et~al., 2019, \mn@doi [\nat] {10.1038/s41586-019-1766-2}, \href
  {https://ui.adsabs.harvard.edu/abs/2019Natur.575..618L} {575, 618}

\bibitem[\protect\citeauthoryear{{Luri} et~al.,}{{Luri} et~al.}{2018}]{luri18}
{Luri} X.,  et~al., 2018, \mn@doi [\aap] {10.1051/0004-6361/201832964}, \href
  {https://ui.adsabs.harvard.edu/abs/2018A&A...616A...9L} {616, A9}

\bibitem[\protect\citeauthoryear{{Maccarone}, {Chomiuk}, {Strader},
  {Miller-Jones}  \& {Sivakoff}}{{Maccarone} et~al.}{2018}]{maccaronengvla}
{Maccarone} T.~J.,  {Chomiuk} L.,  {Strader} J.,  {Miller-Jones} J.,
  {Sivakoff} G.,  2018, {Revealing the Galactic Population of Black Holes}.
p.~711

\bibitem[\protect\citeauthoryear{{Mandel}}{{Mandel}}{2016}]{mandel16}
{Mandel} I.,  2016, \mn@doi [\mnras] {10.1093/mnras/stv2733}, \href
  {https://ui.adsabs.harvard.edu/abs/2016MNRAS.456..578M} {456, 578}

\bibitem[\protect\citeauthoryear{{Mashian} \& {Loeb}}{{Mashian} \&
  {Loeb}}{2017}]{mashian17}
{Mashian} N.,  {Loeb} A.,  2017, \mn@doi [\mnras] {10.1093/mnras/stx1410},
  \href {https://ui.adsabs.harvard.edu/abs/2017MNRAS.470.2611M} {470, 2611}

\bibitem[\protect\citeauthoryear{{Masuda} \& {Hotokezaka}}{{Masuda} \&
  {Hotokezaka}}{2019}]{masuda19}
{Masuda} K.,  {Hotokezaka} K.,  2019, \mn@doi [\apj]
  {10.3847/1538-4357/ab3a4f}, \href
  {https://ui.adsabs.harvard.edu/abs/2019ApJ...883..169M} {883, 169}

\bibitem[\protect\citeauthoryear{{Merloni} et~al.,}{{Merloni}
  et~al.}{2012}]{erosita}
{Merloni} A.,  et~al., 2012, eROSITA Science Book: Mapping the Structure of the
  Energetic Universe, arXiv:1209.3114, \href
  {http://adsabs.harvard.edu/abs/2012arXiv1209.3114M} {}

\bibitem[\protect\citeauthoryear{{Miller-Jones} et~al.,}{{Miller-Jones}
  et~al.}{2021}]{millerjones21}
{Miller-Jones} J. C.~A.,  et~al., 2021, \mn@doi [Science]
  {10.1126/science.abb3363}, \href
  {https://ui.adsabs.harvard.edu/abs/2021Sci...371.1046M} {371, 1046}

\bibitem[\protect\citeauthoryear{{Osuna} et~al.,}{{Osuna} et~al.}{2008}]{adql}
{Osuna} P.,  et~al., 2008, {IVOA Astronomical Data Query Language Version
  2.00}, IVOA Recommendation 30 October 2008 (\mn@eprint {arXiv} {1110.0503}),
  \mn@doi{10.5479/ADS/bib/2008ivoa.spec.1030O}

\bibitem[\protect\citeauthoryear{{Penoyre}, {Belokurov}, {Wyn Evans}, {Everall}
   \& {Koposov}}{{Penoyre} et~al.}{2020}]{penoyre20}
{Penoyre} Z.,  {Belokurov} V.,  {Wyn Evans} N.,  {Everall} A.,   {Koposov}
  S.~E.,  2020, \mn@doi [\mnras] {10.1093/mnras/staa1148}, \href
  {https://ui.adsabs.harvard.edu/abs/2020MNRAS.495..321P} {495, 321}

\bibitem[\protect\citeauthoryear{{Pfahl}, {Rappaport}  \&
  {Podsiadlowski}}{{Pfahl} et~al.}{2003}]{pfahl03}
{Pfahl} E.,  {Rappaport} S.,   {Podsiadlowski} P.,  2003, \mn@doi [\apj]
  {10.1086/378632}, \href
  {https://ui.adsabs.harvard.edu/abs/2003ApJ...597.1036P} {597, 1036}

\bibitem[\protect\citeauthoryear{{Pittard} \& {Dawson}}{{Pittard} \&
  {Dawson}}{2018}]{PIttard18}
{Pittard} J.~M.,  {Dawson} B.,  2018, \mn@doi [\mnras] {10.1093/mnras/sty1025},
  \href {https://ui.adsabs.harvard.edu/abs/2018MNRAS.477.5640P} {477, 5640}

\bibitem[\protect\citeauthoryear{{Pomohaci}, {Oudmaijer}  \&
  {Goodwin}}{{Pomohaci} et~al.}{2019}]{pomohaci19}
{Pomohaci} R.,  {Oudmaijer} R.~D.,   {Goodwin} S.~P.,  2019, \mn@doi [\mnras]
  {10.1093/mnras/stz014}, \href
  {https://ui.adsabs.harvard.edu/abs/2019MNRAS.484..226P} {484, 226}

\bibitem[\protect\citeauthoryear{{Price-Whelan} et~al.,}{{Price-Whelan}
  et~al.}{2020}]{pricewhelan20}
{Price-Whelan} A.~M.,  et~al., 2020, \mn@doi [\apj] {10.3847/1538-4357/ab8acc},
  \href {https://ui.adsabs.harvard.edu/abs/2020ApJ...895....2P} {895, 2}

\bibitem[\protect\citeauthoryear{{Reid}, {McClintock}, {Steiner}, {Steeghs},
  {Remillard}, {Dhawan}  \& {Narayan}}{{Reid} et~al.}{2014}]{reid14}
{Reid} M.~J.,  {McClintock} J.~E.,  {Steiner} J.~F.,  {Steeghs} D.,
  {Remillard} R.~A.,  {Dhawan} V.,   {Narayan} R.,  2014, \mn@doi [\apj]
  {10.1088/0004-637X/796/1/2}, \href
  {http://adsabs.harvard.edu/abs/2014ApJ...796....2R} {796, 2}

\bibitem[\protect\citeauthoryear{{Reig}}{{Reig}}{2011}]{Reig11}
{Reig} P.,  2011, \mn@doi [\apss] {10.1007/s10509-010-0575-8}, \href
  {https://ui.adsabs.harvard.edu/abs/2011Ap&SS.332....1R} {332, 1}

\bibitem[\protect\citeauthoryear{{Reynolds} \& {Miller}}{{Reynolds} \&
  {Miller}}{2011}]{reynoldsmiller11}
{Reynolds} M.~T.,  {Miller} J.~M.,  2011, \mn@doi [\apjl]
  {10.1088/2041-8205/734/1/L17}, \href
  {http://adsabs.harvard.edu/abs/2011ApJ...734L..17R} {734, L17}

\bibitem[\protect\citeauthoryear{{Reynolds}, {Quaintrell}, {Still}, {Roche},
  {Chakrabarty}  \& {Levine}}{{Reynolds} et~al.}{1997}]{reynolds97}
{Reynolds} A.~P.,  {Quaintrell} H.,  {Still} M.~D.,  {Roche} P.,  {Chakrabarty}
  D.,   {Levine} S.~E.,  1997, \mn@doi [\mnras] {10.1093/mnras/288.1.43}, \href
  {https://ui.adsabs.harvard.edu/abs/1997MNRAS.288...43R} {288, 43}

\bibitem[\protect\citeauthoryear{{Rivinius}, {Baade}, {Hadrava}, {Heida}  \&
  {Klement}}{{Rivinius} et~al.}{2020}]{rivinius20}
{Rivinius} T.,  {Baade} D.,  {Hadrava} P.,  {Heida} M.,   {Klement} R.,  2020,
  \mn@doi [\aap] {10.1051/0004-6361/202038020}, \href
  {https://ui.adsabs.harvard.edu/abs/2020A&A...637L...3R} {637, L3}

\bibitem[\protect\citeauthoryear{{Saracino} et~al.,}{{Saracino}
  et~al.}{2021}]{saracino21}
{Saracino} S.,  et~al., 2021, MNRAS in press, \href
  {https://ui.adsabs.harvard.edu/abs/2021arXiv211106506S} {p. arXiv:2111.06506}

\bibitem[\protect\citeauthoryear{{Shahaf}, {Mazeh}, {Faigler}  \&
  {Holl}}{{Shahaf} et~al.}{2019}]{shahaf19}
{Shahaf} S.,  {Mazeh} T.,  {Faigler} S.,   {Holl} B.,  2019, \mn@doi [\mnras]
  {10.1093/mnras/stz1636}, \href
  {https://ui.adsabs.harvard.edu/abs/2019MNRAS.487.5610S} {487, 5610}

\bibitem[\protect\citeauthoryear{{Taylor}}{{Taylor}}{2005}]{topcat}
{Taylor} M.~B.,  2005, in {Shopbell} P.,  {Britton} M.,   {Ebert} R.,  eds,
  Astronomical Society of the Pacific Conference Series Vol. 347, Astronomical
  Data Analysis Software and Systems XIV. p.~29

\bibitem[\protect\citeauthoryear{{Tetarenko}, {Sivakoff}, {Heinke}  \&
  {Gladstone}}{{Tetarenko} et~al.}{2016a}]{watchdog}
{Tetarenko} B.~E.,  {Sivakoff} G.~R.,  {Heinke} C.~O.,   {Gladstone} J.~C.,
  2016a, \mn@doi [\apjs] {10.3847/0067-0049/222/2/15}, \href
  {http://adsabs.harvard.edu/abs/2016ApJS..222...15T} {222, 15}

\bibitem[\protect\citeauthoryear{{Tetarenko} et~al.,}{{Tetarenko}
  et~al.}{2016b}]{tetarenko16}
{Tetarenko} B.~E.,  et~al., 2016b, \mn@doi [\apj] {10.3847/0004-637X/825/1/10},
  \href {https://ui.adsabs.harvard.edu/abs/2016ApJ...825...10T} {825, 10}

\bibitem[\protect\citeauthoryear{{Thompson} et~al.,}{{Thompson}
  et~al.}{2019}]{thompson19}
{Thompson} T.~A.,  et~al., 2019, \mn@doi [Science] {10.1126/science.aau4005},
  \href {https://ui.adsabs.harvard.edu/abs/2019Sci...366..637T} {366, 637}

\bibitem[\protect\citeauthoryear{{Wenger} et~al.,}{{Wenger}
  et~al.}{2000}]{simbad}
{Wenger} M.,  et~al., 2000, \mn@doi [\aaps] {10.1051/aas:2000332}, \href
  {https://ui.adsabs.harvard.edu/abs/2000A&AS..143....9W} {143, 9}

\bibitem[\protect\citeauthoryear{{Wiktorowicz}, {Lu}, {Wyrzykowski}, {Zhang},
  {Liu}, {Justham}  \& {Belczynski}}{{Wiktorowicz}
  et~al.}{2020}]{wiktorowicz20}
{Wiktorowicz} G.,  {Lu} Y.,  {Wyrzykowski} {\L}.,  {Zhang} H.,  {Liu} J.,
  {Justham} S.,   {Belczynski} K.,  2020, \mn@doi [\apj]
  {10.3847/1538-4357/abc699}, \href
  {https://ui.adsabs.harvard.edu/abs/2020ApJ...905..134W} {905, 134}

\bibitem[\protect\citeauthoryear{{Wiktorowicz}, {Middleton}, {Khan}, {Ingram},
  {Gandhi}  \& {Dickinson}}{{Wiktorowicz} et~al.}{2021}]{wiktorowicz21}
{Wiktorowicz} G.,  {Middleton} M.,  {Khan} N.,  {Ingram} A.,  {Gandhi} P.,
  {Dickinson} H.,  2021, \mn@doi [\mnras] {10.1093/mnras/stab2135}, \href
  {https://ui.adsabs.harvard.edu/abs/2021MNRAS.507..374W} {507, 374}

\bibitem[\protect\citeauthoryear{{Yalinewich}, {Beniamini}, {Hotokezaka}  \&
  {Zhu}}{{Yalinewich} et~al.}{2018}]{yalinewich18}
{Yalinewich} A.,  {Beniamini} P.,  {Hotokezaka} K.,   {Zhu} W.,  2018, \mn@doi
  [\mnras] {10.1093/mnras/sty2327}, \href
  {https://ui.adsabs.harvard.edu/abs/2018MNRAS.481..930Y} {481, 930}

\bibitem[\protect\citeauthoryear{{Yamaguchi}, {Kawanaka}, {Bulik}  \&
  {Piran}}{{Yamaguchi} et~al.}{2018}]{yamaguchi18}
{Yamaguchi} M.~S.,  {Kawanaka} N.,  {Bulik} T.,   {Piran} T.,  2018, \mn@doi
  [\apj] {10.3847/1538-4357/aac5ec}, \href
  {https://ui.adsabs.harvard.edu/abs/2018ApJ...861...21Y} {861, 21}

\bibitem[\protect\citeauthoryear{{Yi}, {Sun}  \& {Gu}}{{Yi}
  et~al.}{2019}]{yi19_lamost}
{Yi} T.,  {Sun} M.,   {Gu} W.-M.,  2019, \mn@doi [\apj]
  {10.3847/1538-4357/ab4a75}, \href
  {https://ui.adsabs.harvard.edu/abs/2019ApJ...886...97Y} {886, 97}

\bibitem[\protect\citeauthoryear{{Zurita}, {Casares}  \& {Shahbaz}}{{Zurita}
  et~al.}{2003}]{zurita03}
{Zurita} C.,  {Casares} J.,   {Shahbaz} T.,  2003, \mn@doi [\apj]
  {10.1086/344534}, \href
  {https://ui.adsabs.harvard.edu/abs/2003ApJ...582..369Z} {582, 369}

\makeatother
\end{thebibliography}

%\begin{thebibliography}{99}
%\end{thebibliography}

%%%%%%%%%%%%%%%%%%%%%%%%%%%%%%%%%%%%%%%%%%%%%%%%%%

%%%%%%%%%%%%%%%%% APPENDICES %%%%%%%%%%%%%%%%%%%%%

\newpage
\appendix

\section{Full catalogue}

A few example catalogue entries are listed in Table\,\ref{tab:fullcatalogue}. The full table will be available through CDS\footnote{\url{http://cdsportal.u-strasbg.fr/}}.

\begin{table*}
Primary sample catalogue\\
\begin{tabular}{lcccccccr}
\hline
\hline
RA$_{\rm (EDR3)}$  &  Dec$_{\rm (EDR3)}$  & $G$ & $d$ & \noise & $F_{\rm X}$ & $\Delta$ & Class & Source ID\\
        deg  &  deg           & mag & kpc &   mas    &  10$^{-15}$\,erg\,s$^{-1}$\,cm$^{-2}$ & $''$ & &\\
\hline
% ~39.059799065 &   ~+59.692409783 & 15.20 & 1.58\,$\pm$\,0.11 & 0.36\,$\pm$\,0.03 & 4.4 & 0.4 & -- & --\\
% ~95.685593274 &  ~~~--0.345636239 & 17.47 & 1.50\,$\pm$\,0.24 & 0.37\,$\pm$\,0.17 & 17    & 0.1 & HMXB & 1A\,0620--00\\
% ~53.243901052 & ~~--27.835515446 & 17.46 & 0.35\,$\pm$\,0.02 & 1.22\,$\pm$\,0.02 & 0.11 & 0.3 & Galaxy & Likely incorrect {\tt SIMBAD} class\\
% 254.457544669 &  ~~+35.342357377 & 13.62 & 11.22\,$\pm$\,1.04 & 0.09\,$\pm$\,0.16 & 56450 & 0.7 & LMXB & HZ\,Her\\
~39.059798817&   ~+59.692402581 & 15.20 & 1.43\,$\pm$\,0.11 & 0.36\,$\pm$\,0.01 & 4.4 & 0.4 & -- & --\\
~95.685591321 &  ~~~--0.345659073 & 17.47 & 1.38\,$\pm$\,0.24 & 0.37\,$\pm$\,0.17 & 17    & 0.1 & HMXB & 1A\,0620--00\\
~53.243901052 & ~~--27.835515446 & 17.46 & 0.34\,$\pm$\,0.02 & 1.22\,$\pm$\,0.02 & 0.11 & 0.3 & Galaxy$^\dag$ & 2MASS\,J03325851--2750079\\
254.457539283 &  ~~+35.342322461 & 13.62 & 7.07\,$\pm$\,1.04 & 0.09\,$\pm$\,0.01 & 56450 & 0.7 & LMXB & HZ\,Her\\

\hline

\end{tabular}
\caption{ICRS coordinates from the default EDR3 reference epoch of 2016 are listed. The distance $d$ here is based on parallax inversion and is corrected for zeropoint offset. $F_{\rm X}$ denotes the \chandra\ CSC2 broadband flux (0.5--7\,keV). $\Delta$ denotes the coordinate offset between \gaia\ and \chandra. The \lq Class\rq\ and \lq Source ID\rq\ are those reported by {\tt SIMBAD}, as of Oct 2021. A portion of the catalogue is shown here for reference, with the full catalogue available through CDS. $^\dag$The \lq Galaxy\rq\ in row 3 is likely an incorrect {\tt SIMBAD} class.\label{tab:fullcatalogue}}
\end{table*}

\section{SIMBAD Object Classifications and Assigned Codes}

Tables\,B1 and B2 list the individual source types (the {\tt main\_type} from {\tt SIMBAD}) together with the corresponding assigned short code denoting the broad source category used in the Results, Discussion and some of the figures in the main paper.

We caution that a small fraction of objects ($\sim$\,1\,\%) have unexpected classifications (e.g. extragalactic sources, extended objects such as planetary nebulae [PN], and even candidate planets). Examining the apparent extragalactic sources suggests prior source types are probably in error (either simple transcribing errors between {\tt SIMBAD} and published work, or source confusion). All of them have significant positive parallax measurements consistent with being Galactic objects. The reason that a few PN lie in our primary sample remains unclear; e.g. whether or not the extended nebular emission introduces artificial astrometric uncertainties. Such objects should obviously be treated with caution; but given their small numbers, they will not bias any of our statistical inferences.

\begin{table}
{\scriptsize
Object classifications from SIMBAD\\
\begin{tabular}{r|r}
\hline
\hline
Class  & Assigned Code\\
\hline
                           &\\
                    Unknown&\\
           multiple\_object&\\
                        **&*\\
                      AGB*&*\\
                   BlueSG*&*\\
                        C*&*\\
           Candidate\_AGB*&*\\
            Candidate\_HB*&*\\
            Candidate\_Hsd&*\\
           Candidate\_RGB*&*\\
           Candidate\_RSG*&*\\
             Candidate\_S*&*\\
            Candidate\_SN*&*\\
        Candidate\_brownD*&*\\
      Candidate\_low-mass*&*\\
      Candidate\_post-AGB*&*\\
                       HB*&*\\
                       PM*&*\\
                      Pec*&*\\
                      RGB*&*\\
                    RedSG*&*\\
                        S*&*\\
                      Star&*\\
                   brownD*&*\\
                 low-mass*&*\\
                 post-AGB*&*\\
                       AGN&A\\
            AGN\_Candidate&A\\
                    Assoc*&A\\
                     BYDra&B\\
            Candidate\_CV*&B\\
            Candidate\_EB*&B\\
           Candidate\_HMXB&B\\
            Candidate\_XB*&B\\
                 CataclyV*&B\\
                       EB*&B\\
                      HMXB&B\\
                      LMXB&B\\
                    %LSB\_G&B\\
                      Nova&B\\
                     RSCVn&B\\
              RotV*alf2CVn&B\\
                       SB*&B\\
                Symbiotic*&B\\
                        XB&B\\
                   %BClG&BClG\\
%Blazar\_Candidate&Blazar\_Cand\\
                   %Blue&Blue\\
 BlueStraggler&BlueStraggler\\
             Candidate\_**&C\\
            Candidate\_BSS&C\\
             Candidate\_C*&C\\
           Candidate\_Pec*&C\\
                       Cl*&C\\
                       ClG&C\\
            Compact\_Gr\_G&C\\
                     DkNeb&D\\
                       EmG&E\\
                       FIR&F\\
                        HH&H\\
                       HII&H\\
                        %IG&I\\
                       %ISM&I\\
                %Inexistent&I\\
                       IR&IR\\
                 %LensingEv&L\\
                       MIR&M\\
                    MolCld&M\\
                       NIR&N\\
                     OH/IR&O\\
                     %PairG&P\\
               %PartofCloud&P\\
                    Pulsar&P\\
                       QSO&Q\\
            QSO\_Candidate&Q\\
                 Radio(mm)&R\\
             Radio(sub-mm)&R\\
                    RadioG&R\\
                     RfNeb&R\\
                 Radio&Radio\\
\hline
\end{tabular}
\caption{Object classifications and assigned codes.\label{tab:classes}}
}
\end{table}
\begin{table}
{\scriptsize
Object classifications from SIMBAD\\
\begin{tabular}{r|r}
\hline
\hline
Class  & Assigned Code\\
\hline
                      %SNR?&S\\
                   Seyfert&S\\
                Seyfert\_1&S\\
          Candidate\_TTau*&T\\
                     TTau*&T\\
                       UV&UV\\
        Candidate\_Cepheid&V\\
            Candidate\_LP*&V\\
            Candidate\_Mi*&V\\
          Candidate\_RRLyr&V\\
                   Cepheid&V\\
                  EllipVar&V\\
                Erupt*RCrB&V\\
                 Eruptive*&V\\
                       HV*&V\\
                     HVCld&V\\
             Irregular\_V*&V\\
                      LPV*&V\\
                      Mira&V\\
                 Orion\_V*&V\\
                    PulsV*&V\\
               PulsV*RVTau&V\\
                PulsV*WVir&V\\
              PulsV*delSct&V\\
                     RRLyr&V\\
                     RotV*&V\\
                 Transient&V\\
                        V*&V\\
                       V*?&V\\
                  deltaCep&V\\
                  gammaDor&V\\
                  pulsV*SX&V\\
           Candidate\_WD*&WD\\
                      WD*&WD\\
                      WR*&WD\\
                         X&X\\
            Candidate\_YSO&Y\\
                       YSO&Y\\
                 denseCore&d\\
                     gamma&g\\
                       Ae*&m\\
                       Be*&m\\
            Candidate\_Ae*&m\\
            Candidate\_Be*&m\\
                       Em*&m\\
                     EmObj&m\\
                     BLLac&o\\
                    Galaxy&o\\
                     GinCl&o\\
                  GinGroup&o\\
                   GinPair&o\\
                     GlCl?&o\\
                    GroupG&o\\
                        PN&o\\
                       PN?&o\\
                    Planet&o\\
                   Planet?&o\\
              HotSubdwarf&sd\\
\hline
\end{tabular}
\caption{Continuation of object classifications.\label{tab:classescontd}}
}
\end{table}

%%%%%%%%%%%%%%%%%%%%%%%%%%%%%%%%%%%%%%%%%%%%%%%%%%

% Don't change these lines
\bsp	% typesetting comment
\label{lastpage}
\end{document}